\documentclass{article}

\usepackage{neurips_2023}

\usepackage[utf8]{inputenc} 
\usepackage[T1]{fontenc}    
\usepackage{hyperref}       
\usepackage{url}            
\usepackage{booktabs}       
\usepackage{amsfonts}       
\usepackage{nicefrac}       
\usepackage{microtype}      
\usepackage{xcolor}         

\usepackage{amsmath,amssymb}
\usepackage{mathtools}
\usepackage{amsmath}
\usepackage{graphicx}
\usepackage{algorithm}
\usepackage{algpseudocode}
\usepackage{natbib}
\setcitestyle{authoryear, round, semicolon}

\title{A 3D Conditional Diffusion Model for Image Quality Transfer - An Application to Low-Field MRI}

\author{
  Seunghoi Kim$^{1,2}$\thanks{Corresponding author} \And Henry F. J. Tregidgo$^{1,2}$ \And Ahmed K. Eldaly$^{1,3}$ \And Matteo Figini$^{1,3}$ \And Daniel C. Alexander$^{1,3}$ \\
  \\
  $^1$Centre for Medical Image Computing, University College London, London, UK\\
  $^2$Department of Medical Physics and Biomedical Engineering, University College London,\\ London, UK \\
  $^3$Department of Computer Science, University College London, London, UK
  \\
  \\
  \texttt{\{seunghoi.kim.17, h.tregidgo, a.karam, m.figini, d.alexander\}@ucl.ac.uk}
}

\begin{document}

\maketitle

\begin{abstract}
Low-field (LF) MRI scanners (<1T) are still prevalent in settings with limited resources or unreliable power supply. However, they often yield images with lower spatial resolution and contrast than high-field (HF) scanners. This quality disparity can result in inaccurate clinician interpretations. Image Quality Transfer (IQT) has been developed to enhance the quality of images by learning a mapping function between low and high-quality images. Existing IQT models often fail to restore high-frequency features, leading to blurry output. In this paper, we propose a 3D conditional diffusion model to improve 3D volumetric data, specifically LF MR images. Additionally, we incorporate a cross-batch mechanism into the self-attention and padding of our network, ensuring broader contextual awareness even under small 3D patches. Experiments on the publicly available Human Connectome Project (HCP) dataset for IQT and brain parcellation demonstrate that our model outperforms existing methods both quantitatively and qualitatively. The code is publicly available at \url{https://github.com/edshkim98/DiffusionIQT}.
\end{abstract}

\section{Introduction}
Magnetic Resonance Imaging (MRI) offers detailed, non-invasive medical diagnostics of human anatomy. However, the quality can vary across different scanners, mainly due to the strength of the magnetic field. In resource-limited settings, MRI with lower-strength fields is still common, leading to reduced tissue contrast and signal-to-noise ratio (SNR). To mitigate these challenges, practitioners often acquire images with thick slices, resulting in a reduction in spatial resolution.

\citet{iqt_pio} have introduced a machine learning technique, called image quality transfer (IQT), which aims to enhance the resolution and contrast of low-quality clinical data using rich information in high-quality images. IQT learns a mapping between low and high-quality paired data. Early frameworks \citep{iqt_pio, iqt_harmonization, iqt_uncertainty} demonstrated the effectiveness in enhancing the spatial resolution and information content of diffusion MRI. Subsequent studies \citep{iqt_stochastic, iqt_lisa} proposed decimation simulators to generate low-field MR images to train deep neural networks, aiming to improve contrast and spatial resolution. More recently, to tackle a domain-shift problem in IQT, \citet{synthsr} developed a domain and resolution agnostic model by mapping any input to high-resolution T1-weighted MR images. 
Despite the progress in IQT, significant limitations persist. Current models, especially with large downsampling factors (e.g. x8), tend to produce blurred outputs. Moreover, while patch-based methods have shown satisfactory performance in IQT, as shown by \citet{iqt_stitch}, they tend to create artifacts stemming from limited contextual information and error propagation from the boundaries of patches.

Recently, Generative Adversarial Networks (GANs) have gained popularity in image synthesis tasks, including super-resolution \citep{SRGAN,ESRGAN,sr_soupgan}, and segmentation \citep{Kim2021AGCNAG}. Despite their ability to generate realistic images, GANs are difficult to train due to problems, such as mode collapse and training instability. As an alternative to GANs, diffusion models have recently been developed. They iteratively denoise an image, enabling them to generate high-quality images that surpass GANs in Fréchet inception distance (FID) for image synthesis, as shown in \citet{dmbeatgans}. However, most studies have focused on unconditional settings \citep{ddpm,ddim}, and natural/medical 2D images \citep{srdiff,wu2022medsegdiff}. There have been few studies in 3D domains, especially for volumetric medical data \citep{chung2023solving, 3dpatchsegddpm}. \citet{chung2023solving} adopted diffusion models to solve inverse problems in 3D medical images, but their work was limited to constructing 3D volumes from 2D slices. \citet{3dpatchsegddpm} proposed PatchDDPM for 3D medical image segmentation, utilizing patches during training to reduce memory requirements. However, their approach is constrained by the need for very large patches (e.g., $128^3$ ), which entail significant computational costs.


In this work, we extend the 2D unconditional diffusion model by \citet{variational_diffusion} to a 3D conditional model, abbreviated as DiffusionIQT, to enhance 3D volumetric data, such as low-field MR images. To the best of our knowledge, this is the first work to apply a diffusion model for 3D medical image enhancement. Our proposed network is a 3D neural network, featuring an encoder equipped with transformer and convolution blocks, to capture local and global information. The decoder uses channel-shuffle and convolution blocks to restore fine-detail textures through efficient upsampling. Additionally, we introduce a novel cross-batch mechanism, which aims to share information across patches in a mini-batch through self-attention and padding. This mechanism provides non-local information, leading to improved global consistency and less artifact. Evaluation on the HCP dataset \citep{hcp} for IQT and brain parcellation demonstrates superior performance compared to other existing methods. This highlights the significance of the proposed approach for medical image enhancement.

\section{Methods}
\subsection{Diffusion Process}
Our diffusion process is inspired by \citet{variational_diffusion}, and is composed of a forward process and a reverse process. Assume that we have a clean image $x$, whose noisy version at an arbitrary time step $t$ is $x_{t}$, with $0 \leq t \leq 1$. In the forward process, Gaussian noise is gradually added to a high-quality image until it becomes an isotropic Gaussian noise at $t=1$, where the noise scale is modulated with the cosine scheduler by \citet{diffusion_improved}. 

In the reverse process, the model gradually denoises a high-quality image starting from an isotropic Gaussian noise ($t=1$) until we get a clean high-quality image $x$. However, without an additional prior, it is difficult to control the generation process and predict a target image faithfully. To mitigate this, we add the corresponding low-quality image at each time step as a condition to our network as an input during the reverse process. 
The variational lower bound (vlb) derivation \citep{ddpm} identifies multiple parameterizations of diffusion models, including $\varepsilon$ (noise), $x$-parametrization, and $\nu$ (interpolation between $\varepsilon$ and $x$) by \citet{diffusion_kd}.
Empirically, we found that $x$-parameterization shows superior performance over the others. It comes from the easiness of predicting a deterministic parameter $x$ (target image) compared to the other parametrizations that are non-deterministic. Furthermore, this increases the capacity to model complex transitions, requiring fewer time steps to sample high-quality images. If we denote $\hat{x}_{\theta}$ as a neural network, it takes a noisy image $x_{t}$ with a condition, low-quality image $x_{c}$, as inputs to predict a target image $x$. Hence, the loss function $L$ can then be simply constructed as a difference between $x$ and the output of $\hat{x}_{\theta}$:
\begin{equation}
\label{eq:continuous_loss}
\begin{gathered}
L =  \mathbb{E}_{x,t,\varepsilon}[||x - \hat{x}_{\theta}(\alpha_{t}x + \sigma_{t}\varepsilon,x_{c}, t)||].
\end{gathered}
\end{equation}
where $\alpha_{t}$ and $\sigma_{t}$ are strictly positive scalar-valued functions of $t$, determined by a pre-defined noise scheduler, and $\varepsilon$ is a Gaussian noise. 
\subsection{Network Architecture}
\begin{figure}[H]
\begin{center}    
\includegraphics[scale=0.20]{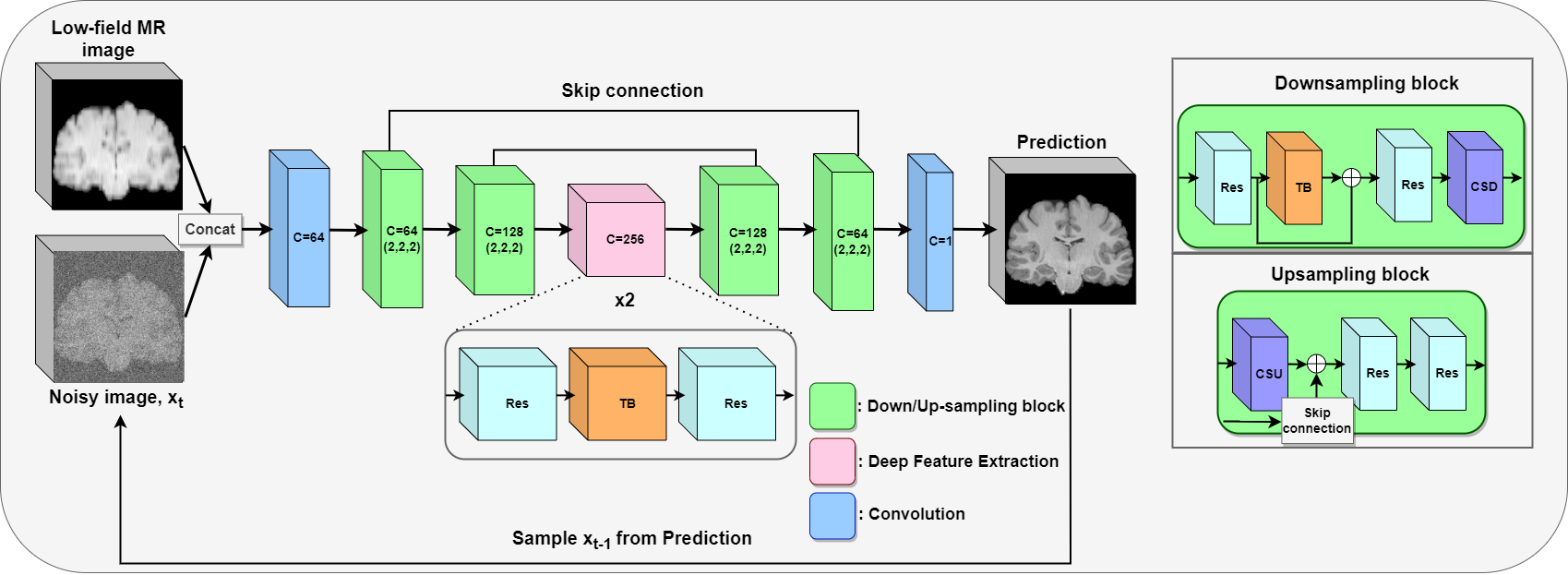}
\caption{Proposed network architecture (C: number of filters, TB: Transformers block, Res: ResNet block, CSD: Channel-shuffle for downsampling, CSU: Channel-shuffle for upsampling). The numbers in parentheses denote downsampling factor in each dimension.}
\label{fig:network}
\end{center}
\end{figure}

As shown in Figure \ref{fig:network}, our novel 3D network, DiffusionIQT, is composed of an encoder and a decoder. It takes concatenated low-field and noisy images as input to predict a target image at each time step. The time embedding is also conditioned on each residual block, in the same manner as \citet{imagen}. The encoder consists of transformers, a 3D extension of \citet{efficient_transformer}, and convolution blocks to introduce long-range dependencies and extract local features. A skip connection is added after the transformer block for an effective fusion of local and global features, and enables vital local spatial information to be preserved, especially crucial in IQT.
With a small input patch size, however, this tends to create boundary artifacts and limits the effectiveness of self-attention. To address these limitations, we introduce a cross-batch mechanism, designed to share information between patches within a mini-batch for self-attention and padding. The mini-batch of 3D patches is constructed from patches that are adjacent and non-overlapping to each other. The key and value in self-attention are computed including the adjacent patches rather than a single patch, mimicking cross-attention. This allows computing attention beyond each patch, enabling features to fuse across the patches. Additionally, we pad patch boundaries with actual values from the adjacent patches before each convolution. These approaches enable each small patch to access broader contextual information, which aids in preserving structural consistency and preventing boundary artifact. Please refer to Appendix \ref{appendix1} for more details. A deep feature extraction module is also added at the bottleneck. This extracts more abstract and high-level features from the input, and it serves to extract a richer representation that compensates for any loss of spatial information.

The decoder consists of an upsampling operation followed by residual blocks. For both downsampling and upsampling feature maps within the network, we employ a 3D channel-shuffle operation, an extension of \citet{espcn}. This method is parameter-free and makes spatial information to be preserved more faithfully than conventional pooling operations. In contrast to the encoder, the decoder does not incorporate transformers. Empirically, we found that transformers can actually deteriorate the model's performance. This may be due to the fact that the long-range dependencies are not pivotal in the decoder, when it comes to reconstructing local fine details and generating high-quality images.
\section{Experiments}
\subsection{Data}
An evaluation was done using the HCP dataset \citep{hcp}. Synthetic low-field MR volumes akin to those obtained using a real scanner were generated using the decimation simulator by \citet{iqt_stochastic}, with a downsampling factor of 8 in the slice direction, mimicking a 4.2mm/1.4mm slice thickness/gap. Of 80 subjects curated randomly, 48 were used for training, 12 for validation, and 20 for testing.
\subsection{Results}
Table \ref{results} presents quantitative comparisons between DiffusionIQT and the four baselines: Interpolation \citep{iqt_pio}, 3D U-Net \citep{iqt_stochastic}, ResU-Net \citep{iqt_stochastic}, and 3D extension of ESRGAN \citep{ESRGAN} in IQT and brain parcellation tasks. Brain parcellation was performed using a pre-trained network from \citet{fastsurfer}. The model was trained using MR images that have same resolution as our evaluated data, which predicts 95 different classes and their statistics in each brain volume. To compare the models quantitatively, we used peak signal-to-noise ratio (PSNR), multi-scale structural similarity index measure (MSSIM), and learned perceptual image patch similarity (LPIPS) for IQT, and mean intersection over union (mIoU) for brain parcellation by using the predicted classes for each voxel. The results demonstrate that DiffusionIQT surpasses all other models by large margins in both tasks. In particular, considerable improvements are observed in the PSNR and LPIPS metrics. This suggests that DiffusionIQT predicts overall voxel intensities and restores high-frequency textures more accurately than the other models. In brain parcellation, our approach outperformed other models by more than 0.1 in mIoU. We can also observe that our model surpasses the performance of other models while having the smallest number of parameters — nearly half the size of 3D ESRGAN. This clearly underscores the effectiveness of the diffusion process and our proposed network architecture.

\begin{figure}[H]
\begin{center}
  \includegraphics[scale=0.28]{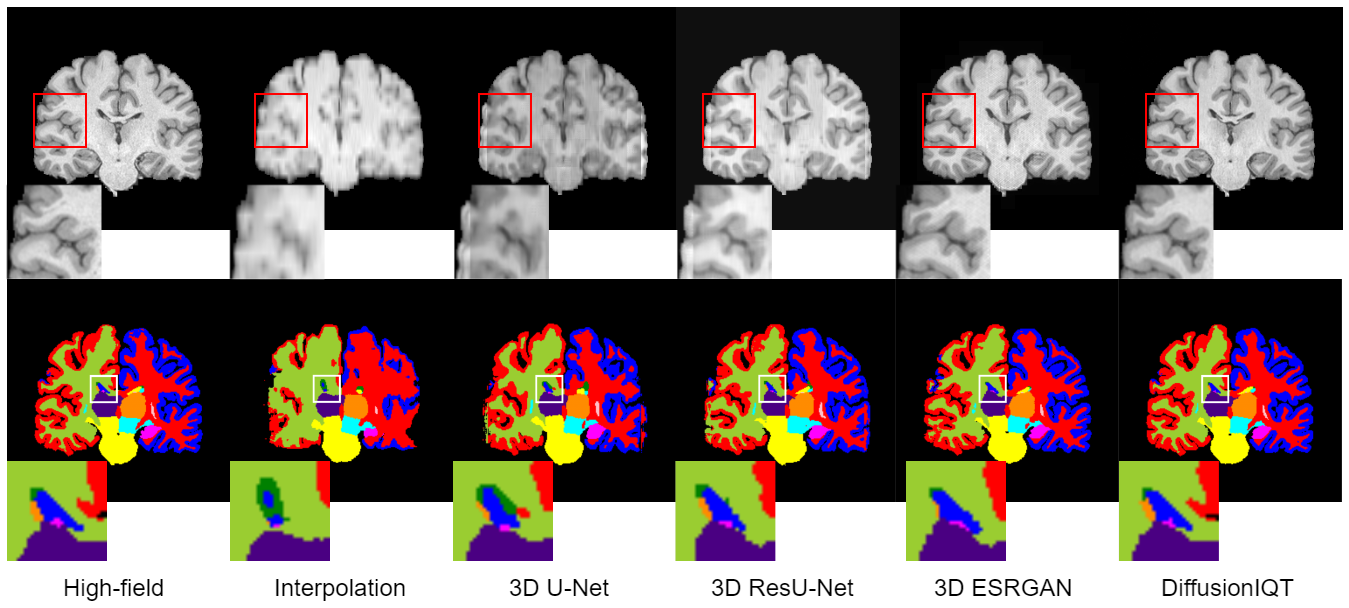}
  \caption{Qualitative comparison on IQT (top) and brain parcellation (bottom)}
  \label{fig:viz}
\end{center}
\end{figure}

Figure \ref{fig:viz} offers qualitative comparisons between the models, illustrating the ability of DiffusionIQT to restore high-frequency textures and accurately delineate tissue structures. As emphasized in the figure, DiffusionIQT more precisely restored the morphology of the sulcus and the lateral ventricle volume compared to other models. While ESRGAN also seemed effective in restoring high-frequency textures, a visible distortion in the size of the restored tissues was observed.
\begin{table}
  \caption{Quantitative comparison results for IQT and brain parcellation. An upward arrow indicates that a higher value is better, and vice versa.}
  \label{results}
  \centering
  \begin{tabular}{cccccc}
    \toprule
     - & Interpolation & 3D U-Net & 3D ResU-Net & 3D ESRGAN & DiffusionIQT \\
    \midrule
    PSNR ($\uparrow$) & 22.4$\pm$0.32  & 26.2$\pm$1.21 & 26.1$\pm$1.45 & 29.7$\pm$1.23 & \textbf{33.7}$\pm$0.80    \\
    MSSIM ($\uparrow$) & 0.805$\pm$0.06 & 0.907$\pm$0.03  & 0.871$\pm$0.07 & 0.940$\pm$0.02 & \textbf{0.968}$\pm$0.02     \\
    LPIPS ($\downarrow$) & 0.293 $\pm$0.018 & 0.232$\pm$0.02 & 0.181$\pm$0.02 & 0.134$\pm$0.01 & \textbf{0.094}$\pm$0.02  \\
    Seg mIoU ($\uparrow$) & 0.488$\pm$0.05 & 0.670$\pm$0.01 & 0.751$\pm$0.008 & 0.749$\pm$0.01 &  \textbf{0.858}$\pm$0.01\\
    Num. Params ($\downarrow$) & - & 48.8M & 37.2M & 28.7M & \textbf{15.2M}  \\
    \bottomrule
  \end{tabular}
\end{table}
\subsection{Module Component Analysis}
In Table \ref{ablation}, we present ablation studies to demonstrate the effectiveness of our proposed modules. The results indicate that the addition of each module significantly enhances overall performance across various image quality metrics. Notably, the deep feature extraction module contributes the most substantial improvement, yielding more than a 10$\%$ increase in all three metrics. While the deep feature extraction module shows much larger gains compared to the cross-batch mechanism, the latter is designed to reduce patch artifacts and enhance structural and contrast consistency, which are subtleties not fully captured by current metrics. Moreover, we achieved this improvement with only a marginal increase in the number of parameters relative to the deep feature extraction module.
\begin{table}
  \caption{Comparative analysis of the proposed modules (x: non-utilized, o: utilized)}
  \label{ablation}
  \centering
  \begin{tabular}{cccccc}
    \toprule
     Deep feature & Cross-batch & PSNR & MSSIM & LPIPS & Num. Params \\
    \midrule
    x & x  & 26.4$\pm$1.41 & 0.848$\pm$0.06 & 0.188$\pm$0.01 & \textbf{9.9M}  \\
    o   & x & 31.5$\pm$1.13 & 0.948$\pm$0.04 & 0.116$\pm$0.01 & 13.6M  \\
    o & o & \textbf{33.7}$\pm$0.80 & \textbf{0.968}$\pm$0.02 & \textbf{0.094}$\pm$0.02 & 15.2M \\
    \bottomrule
  \end{tabular}
\end{table}
\section{Discussion}
Previous work in IQT demonstrated the effective performance with modest downsampling factors. However, these methods often failed to restore high-frequency textures, resulting in blurry predictions. In order to address this limitation, we introduced a 3D conditional diffusion model for IQT to enhance low-field MRI. Our DiffusionIQT is a pioneering work to harness a generative model and 3D vision transformers in tackling IQT challenges. We also proposed cross-batch mechanisms, sidestepping the limitations of learning from small patches. The proposed approach was evaluated using a low-field MRI application in order to recover contrast enhanced images as similar to high field scanners. Comparative experiments using the HCP dataset for IQT and brain parcellation showed that our model consistently outperformed the baseline models, particularly excelling in restoring high-frequency textures and tissue volumes accurately. 

Despite the superior performance in IQT, our work is limited due to the use of synthetic MR images and only healthy subjects. This leaves a future exploration of our model's generalizability to out-of-distribution data and real low-field MR images. 

Nevertheless, we can conclude that leveraging diffusion models signals immense potential in IQT, possibly revolutionizing the medical imaging domain in various clinical scenarios, where diagnostic precision and enhanced image quality are essential.

\section{Acknowledgement}
The authors gratefully acknowledge the helpful input and guidance of the NIHR UCLH Biomedical Research Centre. This work is supported by the EPSRC-funded UCL Centre for Doctoral Training in Intelligent, Integrated Imaging in Healthcare (i4health) EP/S021930/1.
\newpage
\bibliography{References}
\newpage

\appendix
\renewcommand{\thesection}{\Alph{section}}
\renewcommand{\thesubsection}{\thesection.\arabic{subsection}}
\section{Appendix}
\subsection{Network Architecture Details}
\label{appendix1}
In this section, we detail our proposed network architecture, especially the transformers block. 

Our transformers are inspired by \citet{efficient_transformer}. We extend their work to process 3D volumetric data and apply several modifications to effectively capture spatial features. In particular, instead of the traditional method of tokenizing input into non-overlapping patches and then flattening them into 1D, we maintain the 3D shape by tokenizing patches using a 3D depth-wise convolution with both a kernel size and a stride equivalent to the transformer's patch size. Furthermore, we utilize additional depth-wise convolutions for the computation of the query, key, and value. 

\begin{figure}[H]
\centering
\includegraphics[scale=0.3]{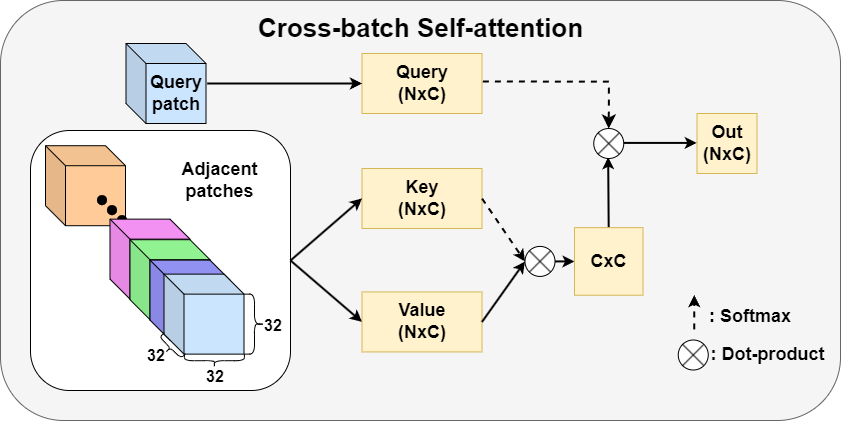}
\caption{Cross-batch mechanism for self-attention. N and C denote the number of patches for self-attention and the number of channels, respectively.}
\label{fig:network_supple}
\end{figure}

As shown in Figure \ref{fig:network_supple}, the self-attention relies on our novel cross-batch mechanism. Unlike conventional vision transformers \citep{vit}, we first perform a dot-product between the key and value. As proposed by \citet{efficient_transformer}, this approach allows for a more efficient operation, as the complexity does not increase with the number of patches. Furthermore, we derive the key and value from adjacent patches, which offers broader contextual information and maximizes the efficacy of self-attention. 

\textbf{Network Architecture} \newline
\begin{itemize}
  \item Input size: $32\times32\times32$
  \item Learning rate: $1e^{-4}$
  \item Number of filters: [64,128,256]
  \item Loss function: Mean Squared Error (MSE)
  \item Activation function: Mish \citep{mish}
  \item Skip-connection scale: $\frac{1}{\sqrt{2}}$
  \item Number of multi-head self-attention (MHSA): 8
  \item Embedding size (C): 512
  \item Number of patches (N): 216
\end{itemize}

\textbf{Diffusion Process} \newline
\begin{itemize}
 \item Number of time steps: 20
 \item Parametrization method: $x$ (target image)
 \item Noise scheduler: Cosine \citep{diffusion_improved}
\end{itemize}

\subsection{Diffusion Process in DiffusionIQT}
In this section, we provide mathematical detail of our diffusion process inspired from \citet{variational_diffusion}.
\subsubsection{Forward Process}
As illustrated in Section 2.1, our diffusion process is mainly composed of a forward process and a reverse process. In the forward process, it involves gradually adding noise to a clean image $x$ to transform it into a noisy image $x_{t}$, where $t$ is an arbitrary time step value $0 \leq t \leq$ 1. Formally, the probability distribution of $x_{t}$ given any arbitrary previous time step $x_{s}$ is defined as:
\begin{equation}
\label{eq:q_sample}
\begin{gathered}
q(x_{t}|x_{s}) = \mathcal{N}(\alpha_{t|s}x_{s}, \sigma_{t|s}^2 I) \\
where\:\:\alpha_{t|s} = \alpha_{t}/\alpha_{s}, and\:\: \sigma_{t|s}^2 = \sigma_{t}^2 - \alpha_{t|s}^2\sigma_{s}^2,
\end{gathered}
\end{equation}
where $\alpha_{t}$ and $\sigma_{t}^2$ denote the diffusion coefficient and noise level at time step $t$, respectively. These parameters are strictly positives and determined by cosine scheduler \citep{diffusion_improved}. \citet{variational_diffusion} chose variance preserving diffusion process, by setting $\alpha_{t}=\sqrt{1-\sigma_{t}^2}$. To sample an arbitrary noisy image $x_{t}$ conditioned on $x$, we sample using $x_{t} = \alpha_{t}x + \sigma_{t}\varepsilon$ where $\varepsilon$ is a Gaussian noise.

\subsubsection{Reverse Process}
In the reverse process, our goal is to iteratively remove the noise from the isotropic Gaussian noise we sampled at time $t=1$, until we obtain a clean image, which is achieved by maximizing the likelihood of $p(x)$. For a finite number of $T$ time steps, by defining $s(i) = (i-1)/T$ and $t(i) = i/T$, the data $x$ can be represented through our generative model as:
\begin{equation}
\label{eq:joint distribution}
\begin{gathered}
p(x) = \int p(x_{1})p(x|x_{0}, x_{c})\prod_{i=1}^{T}p(x_{s(i)}|x_{t(i)}, x_{c}) dx_{0:1},
\end{gathered}
\end{equation}
where $p(x_{1}) = \mathcal{N}(x_{1}; 0,I)$ and $x_{c}$ is a conditioned low-quality MR image. To maximize the likelihood, the neural network is trained to approximate $q(x_{s}|x_{t},x)$ with $p_\theta(x_{s}|x_{t}, x_{c})$, where $\theta$ is learnable parameters. 
Given that both $q(x_{t}|x)$ and $q(x_{s}|x)$ are Gaussian distributions:
\begin{equation}
\label{eq:q_sample2}
\begin{gathered}
q(x_{s}|x_{t}, x) = \frac{q(x_{s},x_{t}|x)}{q(x_{t}|x)} = \frac{q(x_{s}|x)q(x_{t}|x_{s})}{q(x_{t}|x)},
\end{gathered}
\end{equation}
we leverage the conjugate prior property in Bayes' theorem, which ensures that $q(x_{s}|x_{t},x)$ is a Gaussian. Consequently, we model $p_\theta(x_{s}|x_{t}, x_{c})\sim \mathcal{N}(x_{s}; \hat{\mu}_{\theta}(x_{t};t, x_{c}), \sigma^2_{Q}I)$, where $\hat{\mu}_{\theta}$ represents the predicted mean of the posterior term $q(x_{s}|x_{t}, x)$ using a neural network, and $\sigma^2_{Q}$ is the variance in the posterior term, expressed as $\sigma^2_{Q} = \sigma^2_{t|s}\sigma^2_{s}/\sigma^2_{t}$.

Being a Gaussian, this formulation aligns with the derivation of the variational lower bound as defined in \citep{ddpm} for the continuous-time diffusion process. This simplifies our learning objective to computing the mean-squared error as follows:
\begin{equation}
\label{eq:continuous_loss}
\begin{gathered}
L_{t} = \mathbb{E}_{x,t}[||\mu(x_{t},x;t) - \hat{\mu}_{\theta}(x_{t},x_{\theta};t, x_{c})||^2] \\ 
 = \mathbb{E}_{x,t}[||x - \hat{x}_{\theta}||^2] \\
  = \mathbb{E}_{\varepsilon}[||\varepsilon - \hat{\varepsilon}_{\theta}||^2]. \\
\end{gathered}
\end{equation}
Following \citep{variational_diffusion}, the model is parameterized to predict the noise, building upon the approach in \citep{ddpm}. However, in our approach, we parameterize the model to predict the target image $x$ at each time step, which has demonstrated faster convergence in our experiments.

In overall, during training, we randomly sample a time step $t$ from $0\leq t < 1$, to sample a noisy image $x_{t}$. Then, we predict a target image using our neural network $\hat{x}_{\theta}$ given $x_{t}$, $x_{c}$ and $t$. This prediction is then compared with the target image using the reconstruction loss (e.g. L1 or L2) as shown in Equation \ref{eq:continuous_loss}. Once the training is finished, during sampling, given a finite $T$, we discretize time uniformly into T timesteps (segments) and scale them into $[0,1]$. We start from $t = 1$ by sampling isotropic Gaussian noise. After predicting the target image $x$ at time step $t$ given the LF image, we then sample our prediction $\hat{x}$ to obtain the noisy image $x_{t-1}$. This process is iterated until $t=0$.

\end{document}